# Study of a Nano-Square Structure with Defects: Monte Carlo Study

S. Aouini[1], S. Ziti[2], H. Labrim[3] and L. Bahmad[1,*]

[1] Laboratoire de Magnétisme et Physique des Hautes Energies L.M.P.H.E. PPR 13, Mohammed V University of Rabat, Faculty of Sciences, B.P. 1014, Rabat, Morocco.

[2] Laboratoire de Recherche en Informatique, Mohammed V University of Rabat, Faculty of Sciences, B.P. 1014, Rabat, Morocco.

[3] Centre National de l'Energie, des Sciences et des Techniques Nucléaires (CNESTEN), Rabat, Morocco.

**Abstract:**

In this paper, we study the defects effect on magnetic properties of a nano-square structure. This system is composed with alternate layers consisting of the spins $\sigma = 3/2$ and $S = 1$. More precisely, we examine the effect of different physical parameters which are: the coupling exchange interactions, the external magnetic and the crystal fields applied on the studied system. The ground state phase diagrams have been deduced and discussed in different planes involving such different parameters.

Also we study the behavior of this nano-structure by using the Monte Carlo simulations. The behavior of the magnetizations is studied as a function of the external magnetic field, the crystal field, the temperature and the exchange coupling interactions. The magnetizations and the hysteresis cycles are presented and discussed for fixed values of these parameters.

*) Corresponding author: bahmad@fsr.ac.ma (B. L.)

# 1- Introduction :

Nano-materials are the basic elements of nano-science and nanotechnology [1-9]. Nanostructure science and technology is a broad and interdisciplinary area of research and development activity in various scientific domains. That has been growing worldwide in the past few years. It has the potential for revolutionizing the ways in which materials and products are created nature of functionalities that can be accessed. It is already having a significant commercial impact, which will assuredly increase in the future.

The nano- structures having high magnetic properties that important attention such as Cr [10], Co [11], Pt(997) [12], Pt(111) [13], Fe [14] and Fe/Cu(100) [15], these nano-materials are investigated for potential applications in nanotechnology [16,17].

The influence of defects of the non-magnetic has attracted the interest of experimentalists and theoreticians. This contribution of non-magnetic impurities and their usually difficult characterization put also hard constraints, which till now are not always removed. In the last years the number of nominally non-magnetic solids showing magnetic order induced by some kind of defects has increased continuously. The last 10 years investigating the magnetic properties of vacancies and non-magnetic impurities in different oxides, in contrast particularly MgO and ZnO attracted the most interest [18-20]. Using CaO [21] for example, an intensive search is investigated by Elfimov *et al*. [22] for new routes to ferromagnetic nano-materials, it was demonstrated that dilute divalent cation vacancies in oxides with nano-structure lead to a ferromagnetic ground state.

However, Patrykiejew *et al.* [23] has been studied the variation of phase diagrams with the strength of the substrate potential in a lattice gas model for multilayer adsorption. Using transfer matrix and the real spacer normalization group method [24], Benyoussef *et al.* [25] has been studied the layering transitions of a spin-1/2 Ising film. On the other hand, in thin film confined between parallel plates or walls Nakanishi and Fisher [26] have studied the effect of finite size on such transitions. On the other hand, in some of our recent studies [27-30], we have investigated the magnetic properties on the layering transitions in nano-structures systems.

Theoretical investigations and the experimental explorations of the unusual magnetism can be based on model Hamiltonians to study basic physical features of the studies emerged problem [31,32].The numerous methods are explored for the object to examine the magnetic [33] and the physical properties [34] such as effective field theory (EFT) [35], mean field theory (MFT) [36, 37], and Monte Carlo simulations (MCS) [38-42].

The most recent and outstanding works on this field showed that the magnetic properties are not exclusively related to the presence of the magnetic field, but it is depending also of the influence of the interactions between nearest-neighbor atoms [43]. The exchange coupling interactions and the effect of the crystal field are also studied. The focus of this review is to investigate the Monte Carlo method to generate the corresponding phase diagrams. This investigation is made in terms of the magnetization [44] and the hysteresis cycles [45]. A consistent theoretical study of a stable ferromagnetic on a nano-square structure with alternate layers of spins $\sigma = 3/2$ and $S = 1$ consists of several steps. First, the ground state phase diagrams have been presented for the magnetic properties of the corresponding impurities in the studied system. On the other hand, the model and the theoretical formulation used have been presented in this work. Finally, the numerical results found have been discussed and a brief conclusion has been given.

## 2- Model and method:

The studied system is formed with alternate layers($L = 1,2,3,4$) consisting of the spins $\sigma = 3/2$ and $S = 1$. With $L = 1,3$ are the layers of squares of the system occupied by the spins$-\sigma$ and $L = 2,4$ are the layers of squares of the system occupied by the spins$-S$. While, such layer is formed by three thicknesses ($K = 3,6,9,12$), we denote: ($L = \frac{K}{3}$). Our geometry is presented in Fig. 1.

In this section; we examine the magnetic properties of a nano-structure consisting of a ferromagnetic system based on a nano-square structure. In fact, we study the effect of the coupling exchange interactions, the external magnetic and the crystal fields.

The Hamiltonian is described as:

$$\mathcal{H} = -J_{ss}\sum_{i,j} S_i S_j - J_{\sigma\sigma}\sum_{k,l} \sigma_k \sigma_l - J_{s\sigma}\sum_{i,k} S_i \sigma_k - h\sum_i (S_i + \sigma_i) - \Delta\sum_i (S_i^2 + \sigma_i^2) \quad (1)$$

We denote by:

- $J_{\sigma\sigma}$ and $J_{ss}$ are the exchange coupling constant between two first nearest neighbor atoms with spin $\sigma - \sigma$ and $S - S$ respectively.
- $J_{s\sigma}$ is the exchange coupling interaction constant between two nearest neighbor magnetic atoms with spin $S - \sigma$.
- $\Delta$ is the crystal field applied on all S-spins

- h is the external longitudinal magnetic field.

For simplicity, we denote by: $J_{\sigma\sigma} = J_{ss} = J_{s\sigma} = J_{ij}$ which will take a constant value.

The exchange coupling interactions $J_{ij}$ is distributed according to a bimodal law:

$$\mathcal{P}(J_{ij}) = p\delta(J - J_{ij}) + (1-p)\delta(J_{ij}) \quad (2)$$

Where: $\delta$ stands for the Kronecker symbol.

So that the exchange coupling interactions are:

$$J_{ij} = \begin{cases} J & \text{with the probability } p \\ 0 & \text{with the probability } 1-p \end{cases} \quad (3)$$

It is worth to note that for p=0, all sites are diluted. Whereas for p=1, all sites are occupied by a magnetic atom [46].

Using the Metropolis Monte Carlo algorithm, and the Jacknife method [47, 48], we generate new configurations to inspect the magnetic behavior of the studied system. In fact, we perform to examine the equilibrium nano-structure specifically $10^6$ MC steps per spin. By discarding the first $10^5$ generated ones, and by averaging over different initial conditions.

The magnetizations per spin are defined as follows:

$$M_\sigma^L = \frac{1}{N_\sigma^L} \sum_{i=1}^{N_\sigma^L} \sigma_i^L \quad \text{if: } L = 1, 3 \quad (4)$$

$$M_s^L = \frac{1}{N_s^L} \sum_{i=1}^{N_s^L} S_i^L \quad \text{if: } L = 2, 4 \quad (5)$$

With: $N_\sigma^1 = 25$, $N_s^2 = 32$, $N_\sigma^3 = 40$, and $N_s^4 = 48$, spins respectively. And the total number of spins: $N_T = N_\sigma^1 + N_s^2 + N_\sigma^3 + N_s^4 = 25 + 32 + 40 + 48 = 145$.

The susceptibility $\chi$ is defined by:

$$\chi = \beta N(< M^2 > - < M >^2) \quad (6)$$

Where N and M are describing respectively:

$$N = N_\sigma^L + N_s^L \quad (7)$$

$$M = \frac{\sum_{L=1}^{4} N_\sigma^L M_\sigma^L + N_s^L M_s^L}{N} \quad (8)$$

Where: $\beta = 1/k_B T$ with T is the absolute temperature and $k_B$ is Boltzmann's constant. For simplicity we take the value $k_B = 1$.

We define the internal energy per site as follows:

$$E = \frac{1}{N} \langle H \rangle \qquad (9)$$

3- **Results and discussion:**

3-1- **Ground state phase diagrams:**

In this part, we use Monte Carlo simulations for the null value of the temperature (T = 0) to study the magnetic properties. In fact, we calculate the ground state phase diagrams relying of the parameters of the Hamiltonian. By considering a nano-structure composed by four layers (L = 1,2,3,4) of a nano-square with alternate layers containing the spins (3/2,1), with fixed size: $N_\sigma^1 = 25$, $N_s^2 = 32$, $N_\sigma^3 = 40$, and $N_s^4 = 48$. Indeed, the all possible configurations $4 \times 3 \times 4 \times 3 = 144$ are compared and computed in the corresponding ground phase diagrams plotted in Figs. 2(a)-(i).

In the plane (H, Δ) for the fixed values of the exchange coupling interactions: $J_{\sigma\sigma} = J_{ss} = J_{s\sigma} = 1$. A perfect symmetry is found, regarding the external magnetic field H, the all phases are found for the negative values of the crystal field Δ where only two phases are found to be stable for the positive values of this parameter with large regions, see Fig. 2a.

The new phases are appearing in the plane (H, $J_{\sigma\sigma}$) for the fixed values of the exchange coupling interactions: $J_{ss} = J_{s\sigma} = 1$. A perfect symmetry is found, regarding the external magnetic field H in the absence of the crystal field (Δ = 0), see Fig. 2b, the all phases are found for the negative values of the parameter $J_{\sigma\sigma}$ where only two phases are found to be stable for the positive values of this parameter with large regions.

In Fig. 2c, for the fixed values of the exchange coupling interactions: $J_{\sigma\sigma} = J_{ss} = 1$, and in the absence of the crystal field (Δ = 0), we found in the plane (H, $J_{s\sigma}$) the new phases for the negative values of the parameter $J_{s\sigma}$ where only two phases are found to be stable for the positive values of this parameter with a perfect symmetry regarding the external magnetic field H.

In the plane (H, $J_{ss}$) for the fixed values of the exchange coupling interactions: $J_{\sigma\sigma} = J_{s\sigma} = 1$, and in the absence of the crystal field ($\Delta = 0$), four phases are found to be stable for the positive and the negative values of the parameter $J_{ss}$ with deferent regions. The symmetry regarding the external magnetic field H is cracked in this figure, see Fig. 2d.

In Fig. 2e, for the fixed values of the exchange coupling interactions: $J_{ss} = J_{s\sigma} = 1$, and for fixed value of the magnetic field (H = 1), in the plane ($\Delta$, $J_{\sigma\sigma}$) we found some new phases with the large regions when the most of these phases are found to be stable for the negative values of the crystal field $\Delta$.

In the plane ($\Delta$, $J_{s\sigma}$) for the fixed values of the exchange coupling interactions: $J_{ss} = J_{\sigma\sigma} = 1$, and for fixed value of the magnetic field (H = 1), some new phases are found with the large regions when the most of these phases are found to be stable for the negative values of the crystal field $\Delta$, only two phases are found to be stable for the positive values of the parameter $\Delta$. A symmetry regarding the parameter $J_{s\sigma}$ is found in this figure, see Fig. 2f.

In Fig. 2g, for the fixed values of the exchange coupling interactions: $J_{s\sigma} = J_{\sigma\sigma} = 1$, and for fixed value of the magnetic field (H = 1), a symmetry regarding the parameter $J_{ss}$ is found in the plane ($\Delta$, $J_{ss}$) while, we found some new phases with the large regions when the most of these phases are found to be stable for the negative values of the crystal field $\Delta$.

In the plane ($J_{\sigma\sigma}$, $J_{s\sigma}$) for the fixed value of the exchange coupling interaction: $J_{ss} = 1$, and for the fixed value of the magnetic field (H = 1), in the absence of the crystal field ($\Delta = 0$) some new phases are found with the large regions when the most of these phases are found to be stable for the negative values of the parameter $J_{\sigma\sigma}$, while only two phases are found to be stable for the positive values of this parameter.

In the plane ($J_{\sigma\sigma}$, $J_{ss}$) for the fixed value of the exchange coupling interaction: $J_{s\sigma} = 1$, and for the fixed value of the magnetic field (H = 1), in the absence of the crystal field ($\Delta = 0$), two symmetries regarding the parameters $J_{\sigma\sigma}$ and $J_{ss}$ are found in this figure, see Fig. 2i, when some new phases are found to be stable with large regions, only one phase is found to be stable for the negative values of the parameters $J_{\sigma\sigma}$ and $J_{ss}$.

### 3-2- Monte Carlo study:

In this part, we study the effect of the exchange coupling interactions, the external magnetic and the crystal fields. Using Monte Carlo simulations for non-null values of temperature (T ≠ 0) we will study the effect of increasing temperature at which the magnetizations as a function of the external magnetic field, the crystal field and the exchange coupling interactions, respectively. In fact, the corresponding phase diagrams of the magnetizations as a function of the external magnetic field are plotted in Figs. 3(a)-(c). In Fig. 3a, we show the dependence of the hysteresis cycles in the absence of crystal field ($\Delta = 0$), for the fixed values of the exchange coupling interactions: $J_{ss} = J_{\sigma\sigma} = J_{s\sigma} = 1$, and at fixed value of the temperature value. We found a large value of the external magnetic field, when the studied system is named a hard system. By increasing the value of the temperature ($T = 1$), the dependency of the hysteresis cycles is shown, for the fixed values of the exchange coupling interactions: $J_{ss} = J_{\sigma\sigma} = J_{s\sigma} = 1$, and in the absence of crystal field ($\Delta = 0$). When, the value of the external magnetic field is reduced, and the studied system is named a hard system, see Fig. 3b. When the value of the temperature ($T = 2$) increases, the dependency of the hysteresis cycles is shown for the fixed values of the exchange coupling interactions: $J_{ss} = J_{\sigma\sigma} = J_{s\sigma} = 1$, and in the absence of crystal field ($\Delta = 0$), the value of the external magnetic field is reduced more and more, when the system is named a soft system, see Fig. 3c.

The corresponding phase diagrams of the magnetizations as a function of the crystal field $\Delta$ are plotted in Figs. 4(a)-(c). Indeed, at fixed value of the temperature ($T = 0.5$), for the fixed values of the exchange coupling interactions: $J_{ss} = J_{\sigma\sigma} = J_{s\sigma} = 1$, and for fixed value of the external magnetic field ($H = 1$), see Fig. 4a, the behavior of the magnetizations of each two layers of the same spins ($m_1$, $m_3$ and $m_2$, $m_4$) is of first order transition type from the negative values to the positive values of the magnetizations. With increasing the value of the temperature ($T = 1$), for the fixed values of the exchange coupling interactions: $J_{ss} = J_{\sigma\sigma} = J_{s\sigma} = 1$, and for fixed value of the external magnetic field ($H = 1$), the behavior of the magnetizations of each two layers of the same spins stay of first order transition type starting of the negative values to the positive values of the magnetizations, see Fig. 4b. But when increasing the value of the temperature ($T = 2$), for the fixed values of the exchange coupling interactions: $J_{ss} = J_{\sigma\sigma} = J_{s\sigma} = 1$, and for fixed value of the external magnetic field ($H = 1$), the behavior of the magnetizations of each two layers of the same spins become the second order transition type starting of the negative values to the positive values of the magnetizations, see Fig. 4c.

The corresponding phase diagrams of the magnetizations as a function of the exchange coupling interactions are plotted in Figs. 5(a)-(i). In fact, for the fixed value of the temperature ($T = 0.5$), we plot in Fig. 5a the corresponding phase diagrams of the magnetizations as a function of the parameter $J_{\sigma\sigma}$, for the fixed values of the exchange coupling interactions: $J_{ss} = J_{s\sigma} = 1$, and for fixed value of the external magnetic field ($H = 1$), in the absence of the crystal field ($\Delta = 0$) each two layers of the same spins ($m_1$, $m_3$ and $m_2$, $m_4$) behave the same way. The layers $m_1$ and $m_2$ transit simultaneously the first order transition type starting of the negative values of the parameter and stabilize for the positive values of this parameter. When, the layers $m_3$ and $m_4$ keep the constant value for the negative and the positive values of the parameter $J_{\sigma\sigma}$. By increasing the value of the temperature ($T = 1$), for the fixed values of the exchange coupling interactions: $J_{ss} = J_{s\sigma} = 1$, and for fixed value of the external magnetic field ($H = 1$), in the absence of the crystal field ($\Delta = 0$) each two layers of the same spins ($m_1$, $m_3$ and $m_2$, $m_4$) behave the same way. The layers $m_1$ and $m_2$ transit simultaneously the second order transition type from the negative values of the parameter and stabilize for the positive values of this parameter. When, the layers $m_3$ and $m_4$ keep the constant value for the negative and the positive values of the parameter $J_{\sigma\sigma}$, see Fig. 5b. With increasing the temperature ($T = 2$), for the fixed values of the exchange coupling interactions: $J_{ss} = J_{s\sigma} = 1$, and for fixed value of the external magnetic field ($H = 1$), in the absence of the crystal field ($\Delta = 0$) the layers $m_1$, $m_2$ and $m_3$ transit for the negative values of the magnetizations, and the layer $m_4$ transit for the positive values of the magnetizations. Finally, all the layers stabilize for the positive values of the parameter $J_{\sigma\sigma} \geq 2$, see Fig. 5c.

Indeed, for the fixed value of the temperature ($T = 0.5$), we plot in Fig. 5d the corresponding phase diagrams of the magnetizations as a function of the parameter $J_{s\sigma}$, for the fixed values of the exchange coupling interactions: $J_{ss} = J_{\sigma\sigma} = 1$, and for fixed value of the external magnetic field ($H = 1$), in the absence of the crystal field ($\Delta = 0$) the layers $m_1$ and $m_2$ transit for the negative values of the magnetizations, and the layers $m_3$ and $m_4$ transit for the positive values of the magnetizations. While, all the layers stabilize for the positive values of the parameter $J_{\sigma\sigma} \geq 1.5$. By increasing the value of the temperature ($T = 1$), for the fixed values of the exchange coupling interactions: $J_{ss} = J_{\sigma\sigma} = 1$, and for fixed value of the external magnetic field ($H = 1$), in the absence of the crystal field ($\Delta = 0$) the layers $m_1$ and $m_2$ transit simultaneously the first order transition type from the negative values of the magnetizations and stabilize for the positive values of the magnetizations. When, the layers $m_3$ and $m_4$ keep the constant value for the positive values of the magnetizations, see Fig. 5e. With increasing the temperature ($T = 2$), for the fixed values of the exchange coupling interactions: $J_{ss} = J_{s\sigma} = 1$, and for fixed value of the external magnetic field ($H = 1$), in the absence of the crystal field ($\Delta = 0$) the behavior of the magnetizations is deferent for each layer. The layer $m_1$ transits the second order transition type from the negative value of the magnetizations and stabilize for the positive value of the magnetizations, the layer $m_2$ transits the first order transition type from the negative value of the magnetizations and stabilize for the positive value of the magnetizations, the layer $m_3$ stays stable for the positive value of the magnetizations, and the layer $m_4$ transits for the positive value of the magnetizations with a downfall until the negative value of the magnetizations. Finally, all the layers $m_1$, $m_2$, $m_3$ and $m_4$ stabilize for the positive values of the parameter $J_{s\sigma} \geq 1$, see Fig. 5f.

In fact, for the fixed value of the temperature (T = 0.5), we plot in Fig. 5g the corresponding phase diagrams of the magnetizations as a function of the parameter $J_{ss}$, for the fixed values of the exchange coupling interactions: $J_{s\sigma} = J_{\sigma\sigma} = 1$, and for fixed value of the external magnetic field (H = 1), in the absence of the crystal field ($\Delta = 0$) the layer $m_1$ keeps the constant value for the positive value of the magnetizations. When, the layers $m_2$, $m_3$ and $m_4$ transit simultaneously the second order transition type and stabilize for the positive values of the magnetizations. With increasing the temperature (T = 1), for the fixed values of the exchange coupling interactions: $J_{s\sigma} = J_{\sigma\sigma} = 1$, and for fixed value of the external magnetic field (H = 1), in the absence of the crystal field ($\Delta = 0$) the layer $m_1$ keeps the constant value for the positive value of the magnetizations, the layer $m_2$ transits the second order transition type for the positive values of the magnetizations, and the layers $m_3$ and $m_4$ transit simultaneously the second order transition type from the null value of the magnetizations to the positive values of the magnetizations. While, all the layers stabilize for the positive values of the parameter $J_{ss} \geq 0.5$, see Fig. 5h. The same scenario is repeated by increasing the value of the temperature (T = 2), for the fixed values of the exchange coupling interactions: $J_{s\sigma} = J_{\sigma\sigma} = 1$, and for fixed value of the external magnetic field (H = 1), in the absence of the crystal field ($\Delta = 0$) the layers $m_1$ and $m_2$ transit the second order transition type for the positive values of the magnetizations, and the layers $m_3$ and $m_4$ transit simultaneously the second order transition type from the null value of the magnetizations to the positive values of the magnetizations. While, all the layers stabilize for the positive values of the parameter $J_{ss} \geq 1.5$, see Fig. 5i.

### 4- Conclusion:

In this paper, we have studied the defects effect on magnetic properties of a nano-square structure with alternate layers consisting of a ferromagnetic system. The effect of the coupling exchange interactions, the external magnetic and the crystal fields is also investigated in this study. The ground state phase diagrams in different planes were presented in our result for different values of the phase space parameters. The majority stable configurations are found for the negative values of these parameters. On the other hand, it is found that a perfect symmetry appears with respect to the external magnetic field H.

The behavior of the magnetizations and the hysteresis loops have been analyzed in several phase diagrams by using the Monte Carlo simulations. When increasing the temperature values the behavior of the magnetizations has been shifted from the first to the second transition type. The behavior of the hysteresis cycles has been deduced, showing the change of the studied system from the hard system to the soft one, in different phase diagrams.

**Figures captions**

**Fig.1:** A geometry of the studied system composed by four nano-square magnetic layers with alternate spins $\sigma = 3/2$ and $S = 1$, containing $N_\sigma^1 = 25$, $N_S^2 = 32$, $N_\sigma^3 = 40$, and $N_S^4 = 48$.

**Fig. 2:** Ground state phase diagrams of the studied system:

(a) in the plane $(H, \Delta)$ for $J_{\sigma\sigma} = J_{ss} = J_{s\sigma} = 1$.

(b) in the plane $(H, J_{\sigma\sigma})$ for $J_{ss} = J_{s\sigma} = 1$ and $\Delta = 0$.

(c) in the plane $(H, J_{s\sigma})$ for $J_{\sigma\sigma} = J_{ss} = 1$ and $\Delta = 0$.

(d) in the plane $(H, J_{ss})$ for $J_{\sigma\sigma} = J_{s\sigma} = 1$ and $\Delta = 0$.

(e) in the plane $(\Delta, J_{\sigma\sigma})$ for $J_{s\sigma} = J_{ss} = 1$ and $H = 0$.

(f) in the plane $(\Delta, J_{s\sigma})$ for $J_{\sigma\sigma} = J_{ss} = 1$ and $H = 0$.

(g) in the plane $(\Delta, J_{ss})$ for $J_{\sigma\sigma} = J_{s\sigma} = 1$ and $H = 0$.

(h) in the plane $(J_{\sigma\sigma}, J_{s\sigma})$ for $J_{ss} = 1$, $H = 1$ and $\Delta = 0$.

(i) in the plane $(J_{\sigma\sigma}, J_{ss})$ for $J_{s\sigma} = 1$, $H = 1$ and $\Delta = 0$.

**Fig. 3:** Magnetization profiles:

(a) Magnetizations as a function of H, for $J_{\sigma\sigma} = J_{ss} = J_{s\sigma} = 1$, and $\Delta = 0$ for $T = 0.5$.

(b) Magnetizations as a function of H, for $J_{\sigma\sigma} = J_{ss} = J_{s\sigma} = 1$, and $\Delta = 0$ for $T = 1.0$.

(c) Magnetizations as a function of H, for $J_{\sigma\sigma} = J_{ss} = J_{s\sigma} = 1$, and $\Delta = 0$ for $T = 2.0$.

**Fig. 4:** Magnetization profiles:

(a) Magnetizations as a function of $\Delta$, for $J_{\sigma\sigma} = J_{ss} = J_{s\sigma} = 1$, and $H = 1$ for $T = 0.5$.

(b) Magnetizations as a function of $\Delta$, for $J_{\sigma\sigma} = J_{ss} = J_{s\sigma} = 1$, and $H = 1$ for $T = 1.0$.

(c) Magnetizations as a function of $\Delta$, for $J_{\sigma\sigma} = J_{ss} = J_{s\sigma} = 1$, and $H = 1$ for $T = 2.0$.

**Fig. 5:** Magnetization profiles:

(a) Magnetizations as a function of $J_{\sigma\sigma}$, for $J_{ss} = J_{s\sigma} = 1$, $H = 1$ and $\Delta = 0$ for $T = 0.5$.

(b) Magnetizations as a function of $J_{\sigma\sigma}$, for $J_{ss} = J_{s\sigma} = 1$, $H = 1$ and $\Delta = 0$ for $T = 1.0$.

(c) Magnetizations as a function of $J_{\sigma\sigma}$, for $J_{ss} = J_{s\sigma} = 1$, $H = 1$ and $\Delta = 0$ for $T = 2.0$.

(d) Magnetizations as a function of $J_{s\sigma}$, for $J_{\sigma\sigma} = J_{ss} = 1$, $H = 1$ and $\Delta = 0$ for $T = 0.5$.

(e) Magnetizations as a function of $J_{s\sigma}$, for $J_{\sigma\sigma} = J_{ss} = 1$, $H = 1$ and $\Delta = 0$ for $T = 1.0$.

(f) Magnetizations as a function of $J_{s\sigma}$, for $J_{\sigma\sigma} = J_{ss} = 1$, $H = 1$ and $\Delta = 0$ for $T = 2.0$.

(g) Magnetizations as a function of $J_{ss}$, for $J_{\sigma\sigma} = J_{s\sigma} = 1$, $H = 1$ and $\Delta = 0$ for $T = 0.5$.

(h) Magnetizations as a function of $J_{ss}$, for $J_{\sigma\sigma} = J_{s\sigma} = 1$, $H = 1$ and $\Delta = 0$ for $T = 1.0$.

(i) Magnetizations as a function of $J_{ss}$, for $J_{\sigma\sigma} = J_{s\sigma} = 1$, $H = 1$ and $\Delta = 0$ for $T = 2.0$.

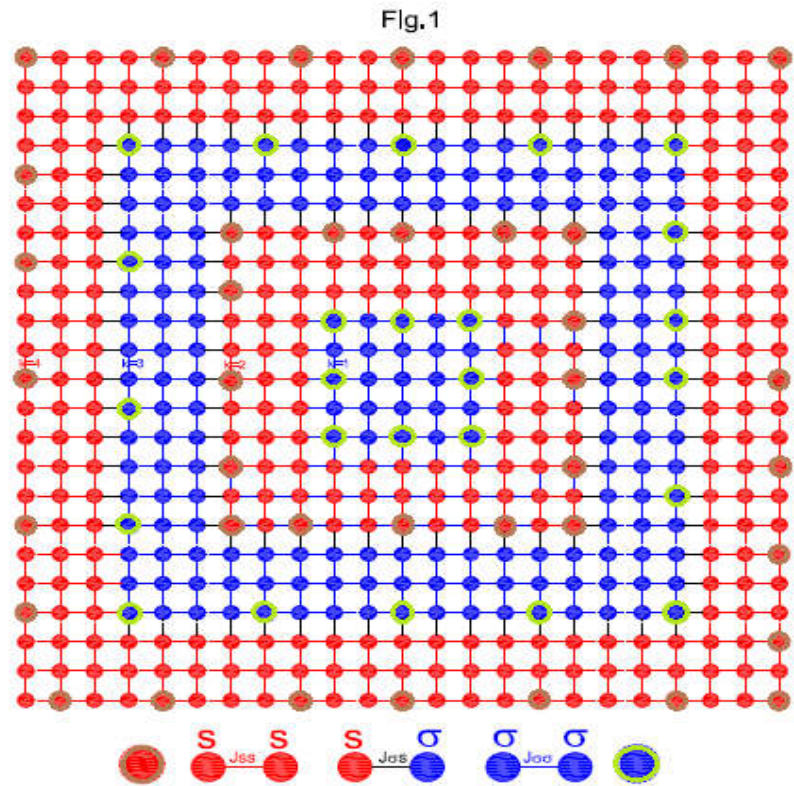

Fig.1

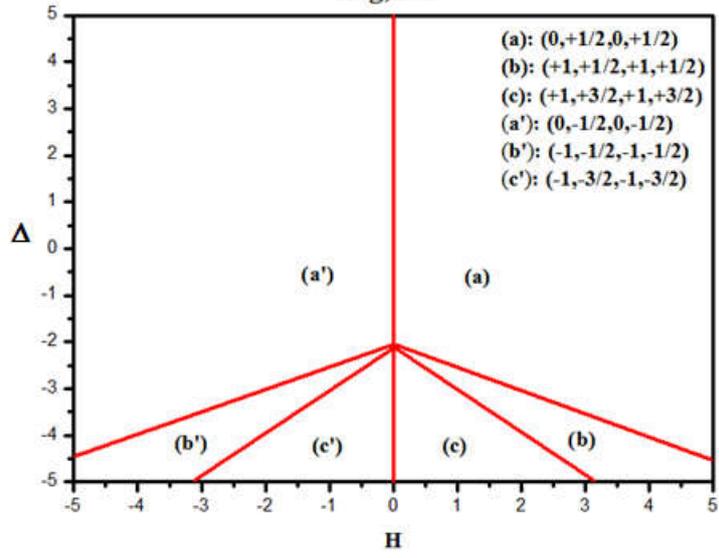

Fig, 2 a

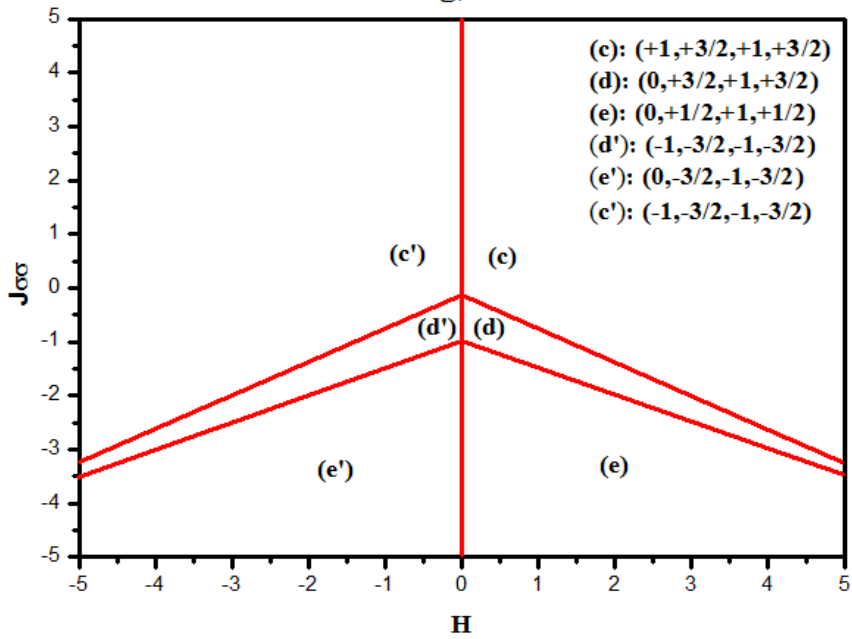

Fig, 2 b

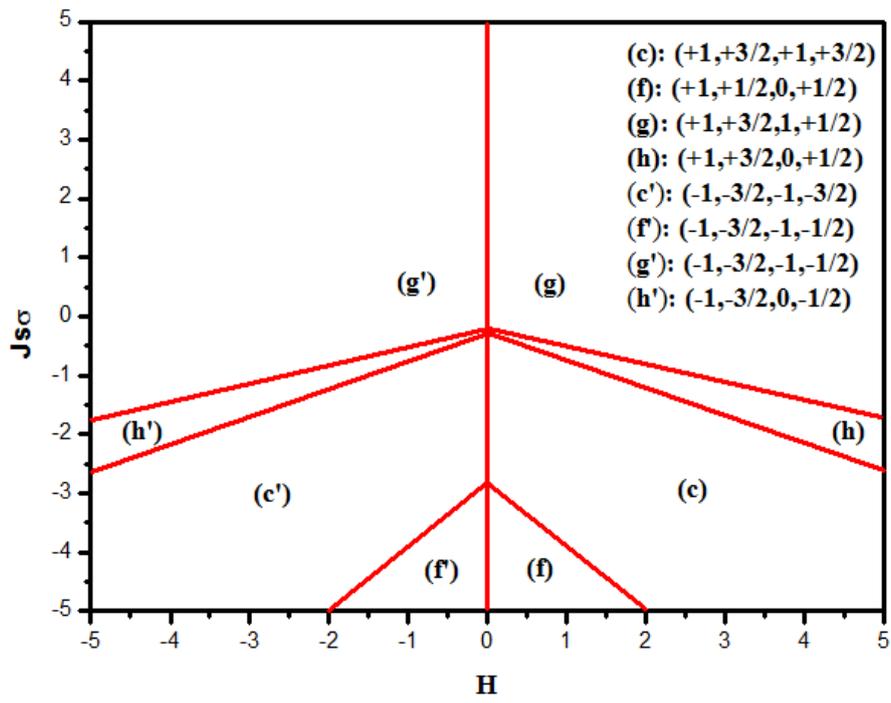

Fig. 2 c

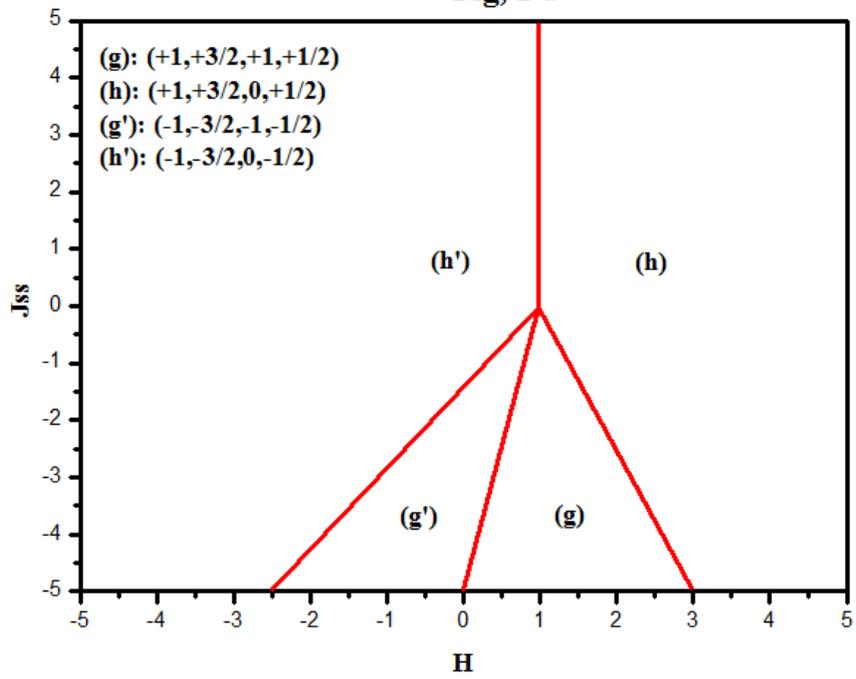

Fig. 2 d

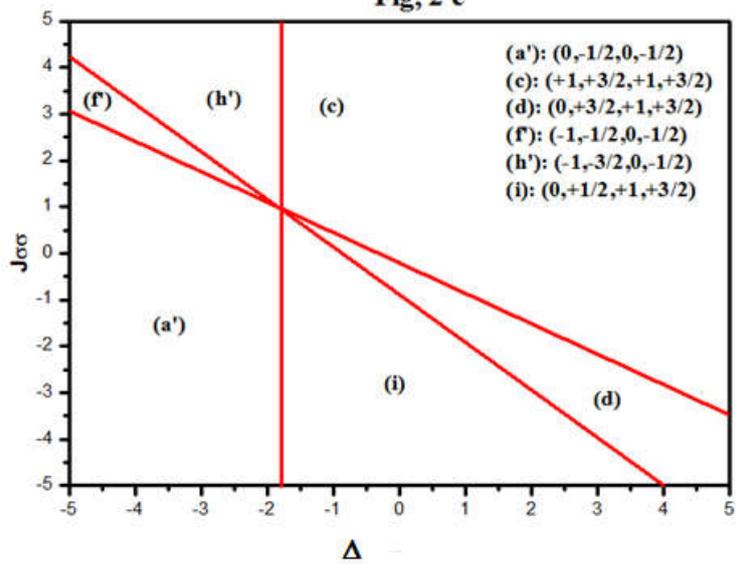

Fig. 2 e

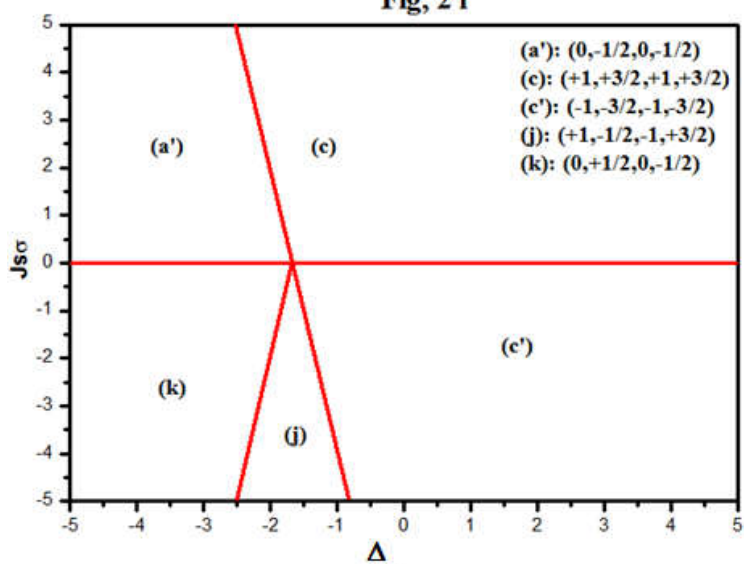

Fig. 2 f

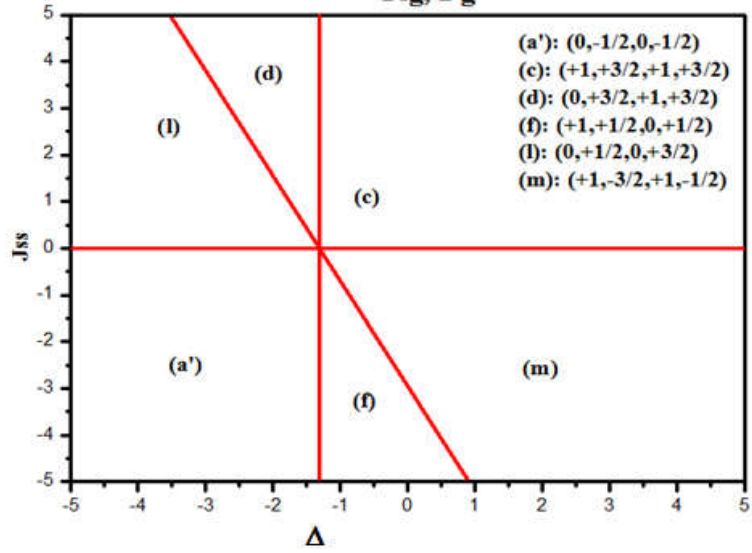

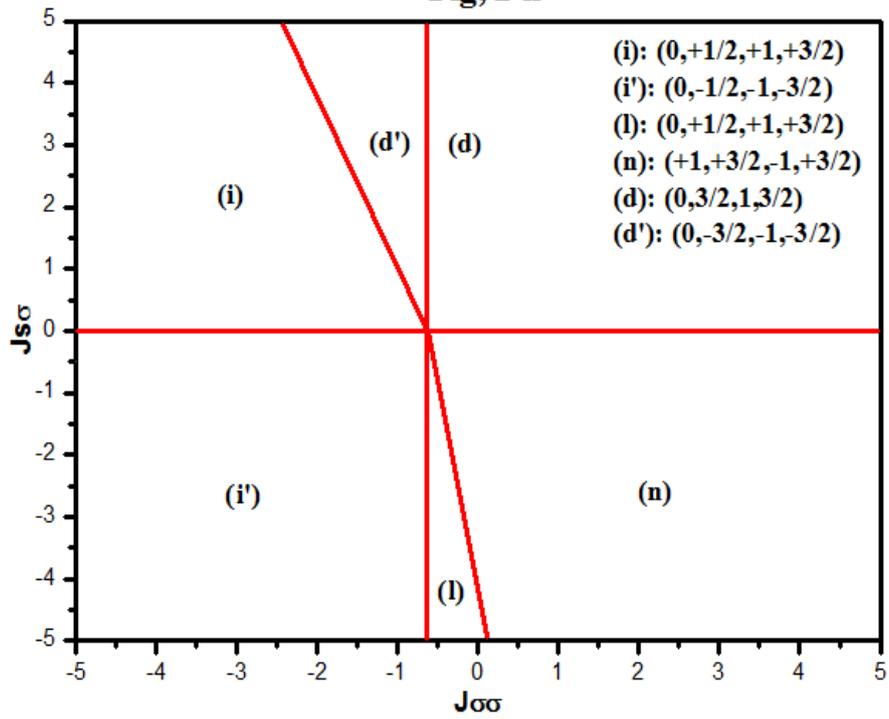

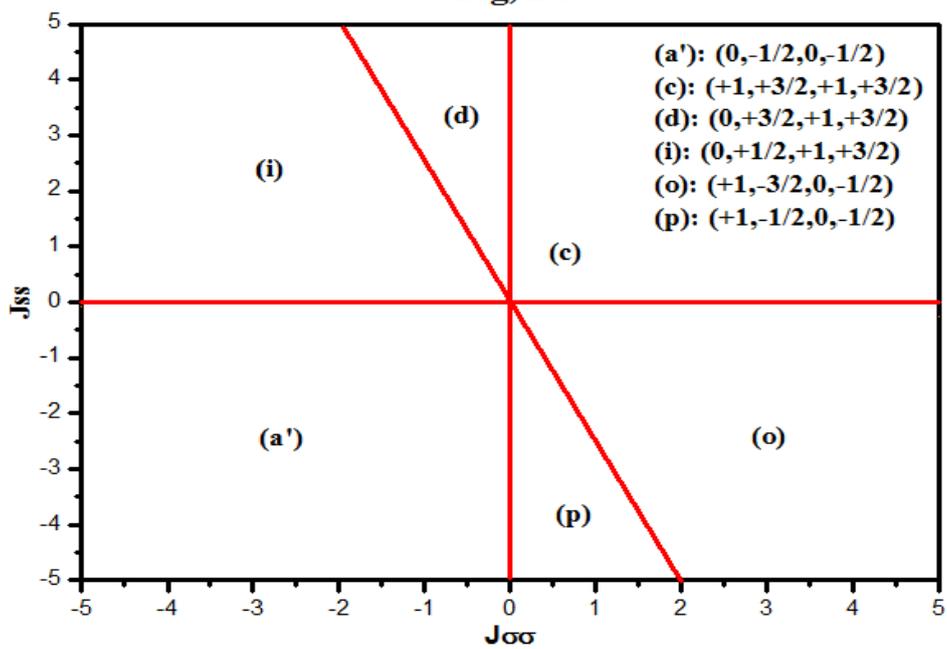

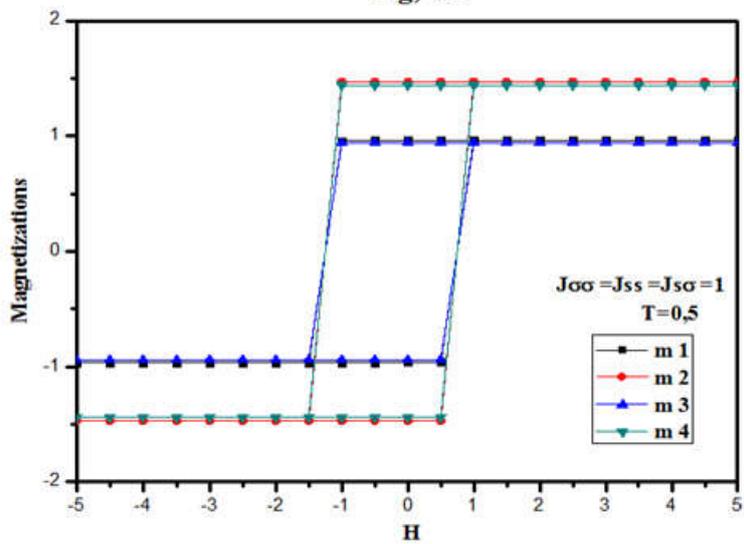

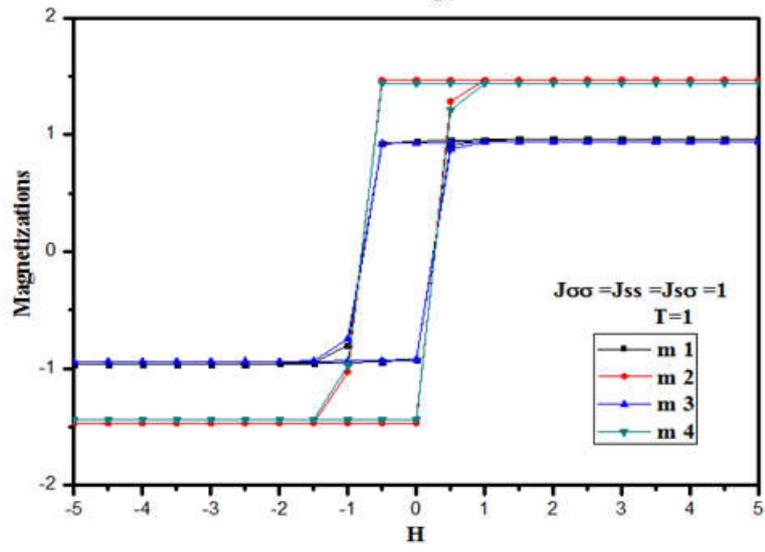

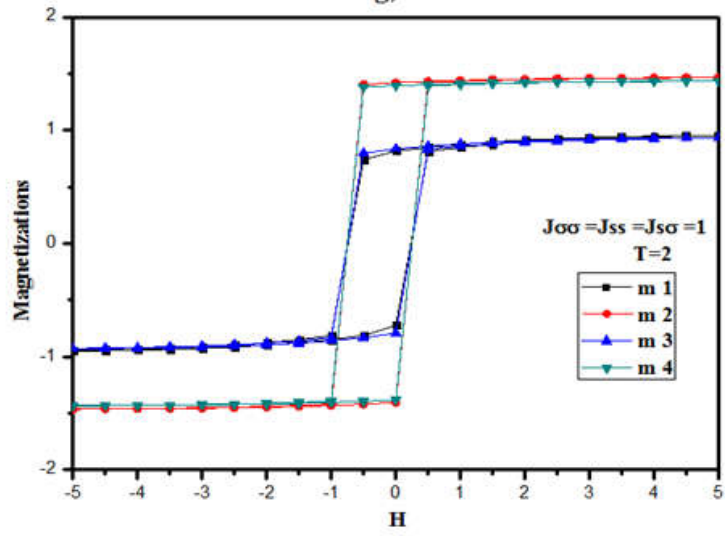

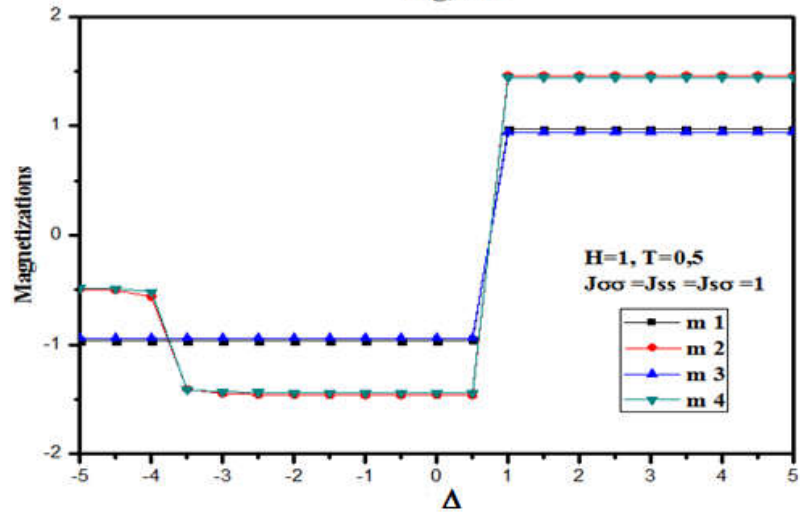

Fig, 4a

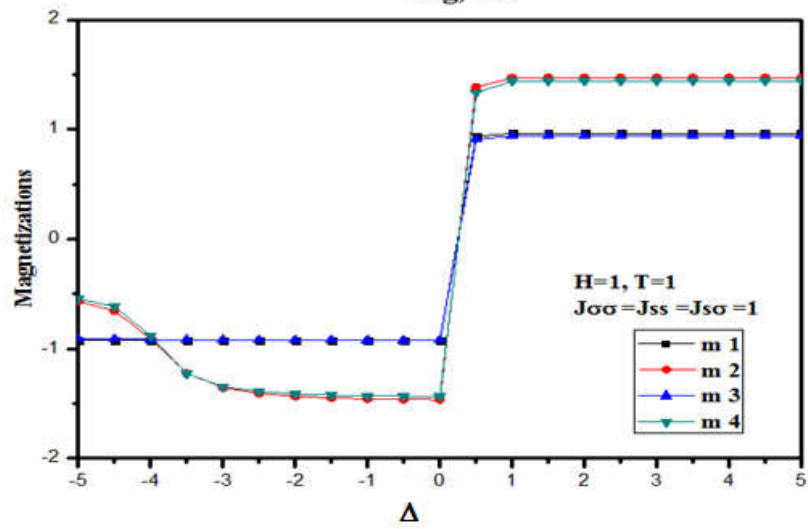

Fig, 4b

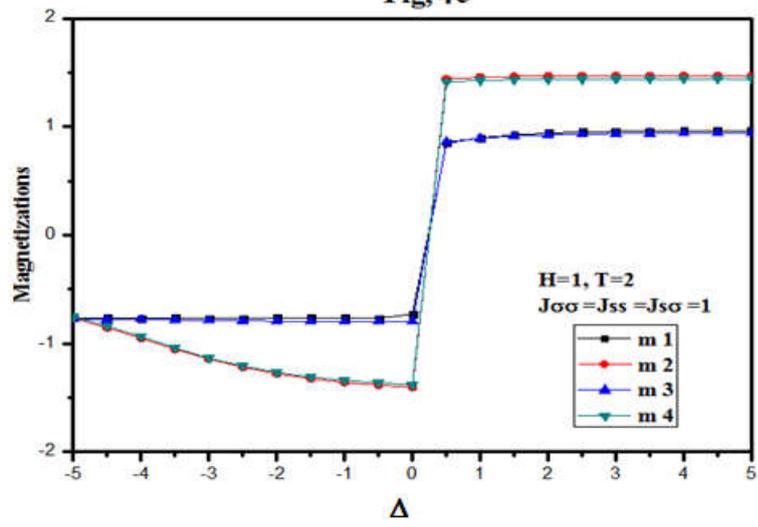

Fig. 4c

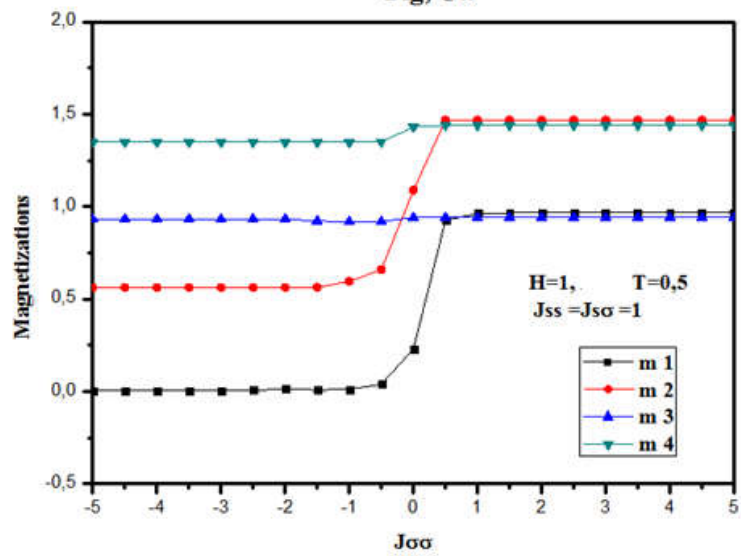

Fig. 5a

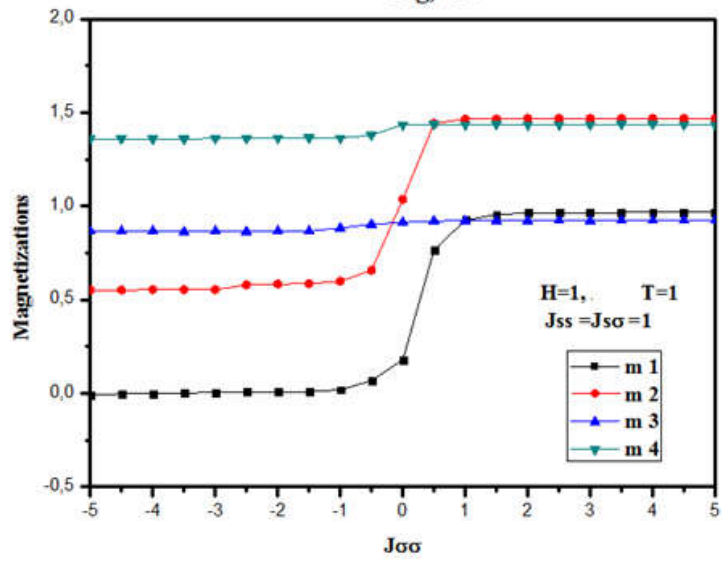

Fig, 5b

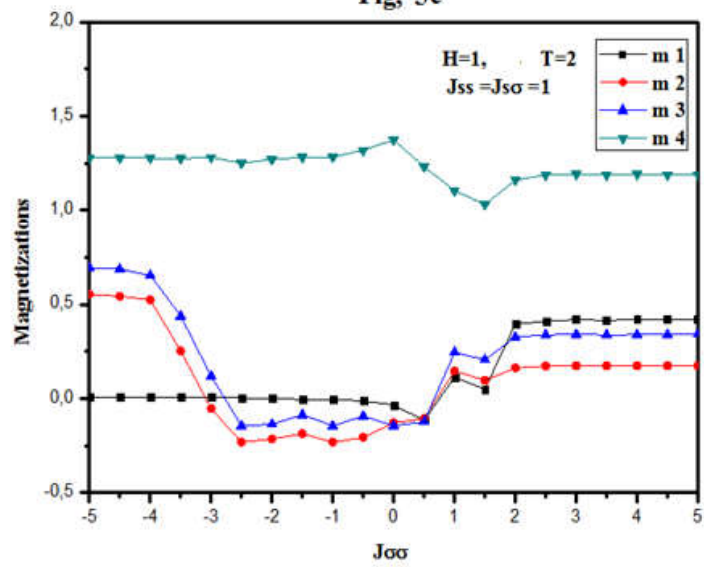

Fig, 5c

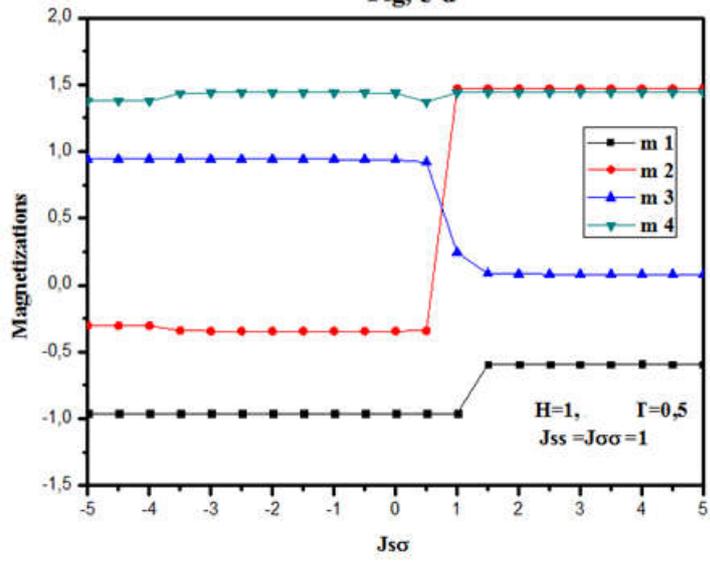

Fig, 5 d

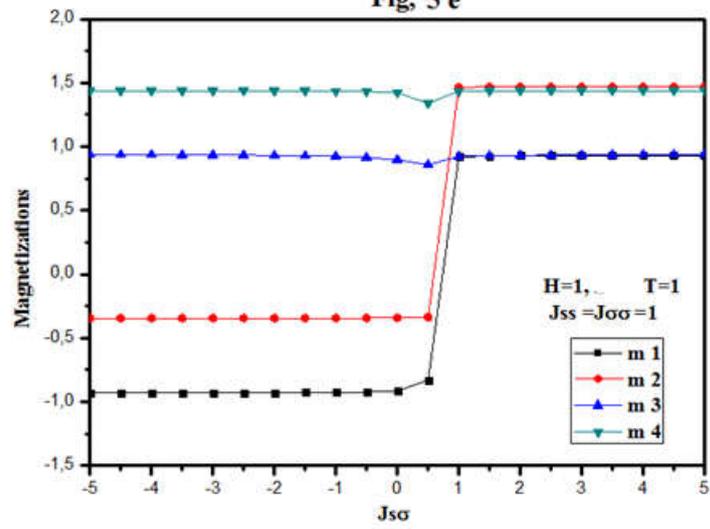

Fig, 5 e

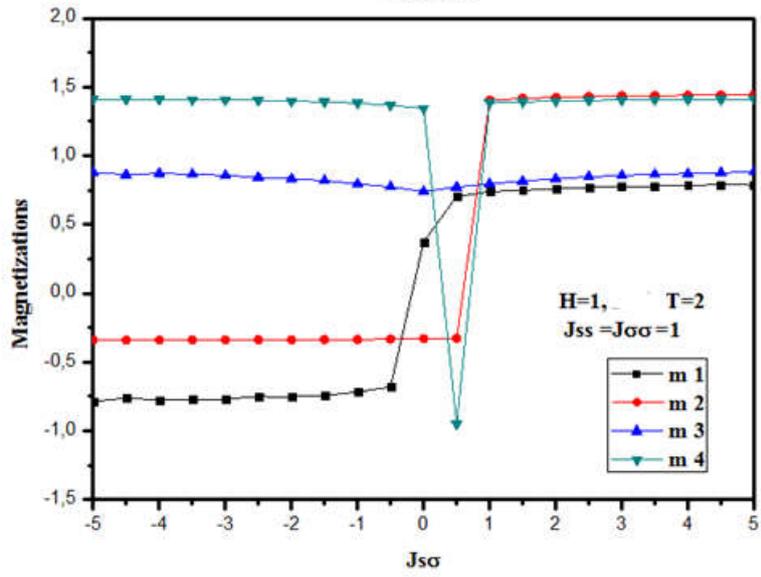

Fig, 5 f

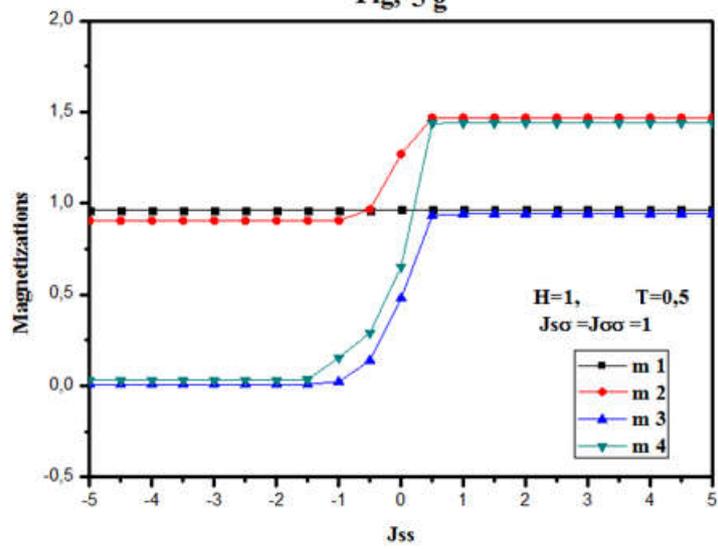

Fig, 5 g

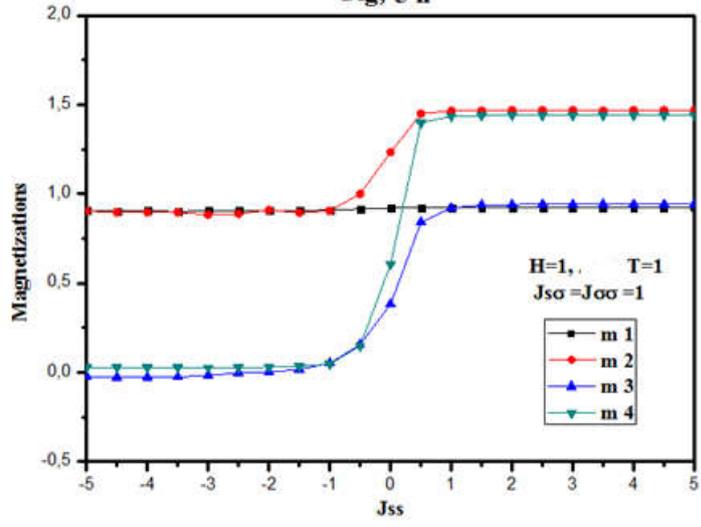

Fig, 5 h

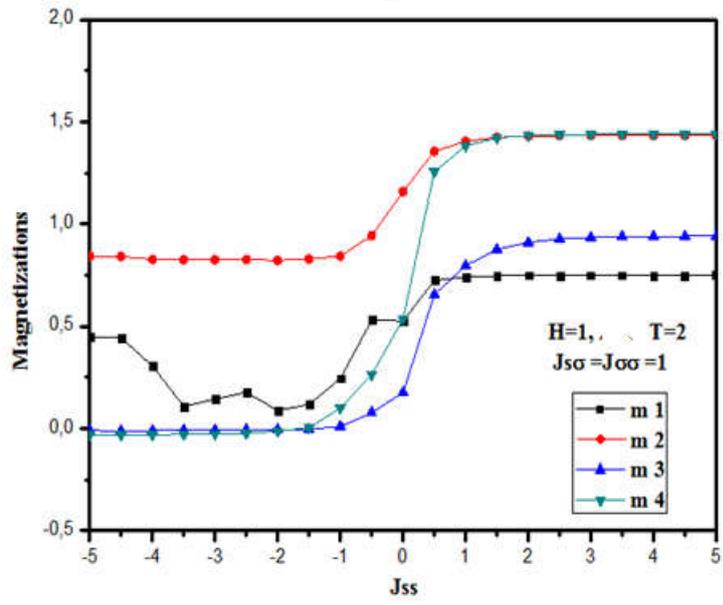

Fig, 5 i